\documentclass{sf2a-conf2025}
\usepackage{graphicx}
\usepackage{hyperref}
\usepackage[]{natbib}
\usepackage{epstopdf}
\usepackage{xcolor}

\begin{document}

\TitreGlobal{SF2A 2025}

\title{A unique solution to overcome the barriers to planetesimal formation at low dust-to-gas ratio}

\runningtitle{Overcoming the barriers to planetesimal formation}

\author{H. Meheut}\address{Université Côte d’Azur, Observatoire de la Côte d’Azur, CNRS, Laboratoire Lagrange, Nice, France.}

\author{F. A. Gerosa$^{1,}$}\address{Mullard Space Science Laboratory, University College London, Holmbury St. Mary, Dorking, UK}

\author{J. Bec}\address{Université Côte d’Azur, CNRS, Institut de Physique de Nice, France}

\setcounter{page}{237}


\maketitle


\begin{abstract}
In the incremental growth model, planetesimal formation constitutes the least understood step in the process of planetary formation. The two main difficulties in this regard are the collision/fragmentation and the drift barriers. Numerous solutions have been proposed to overcome these barriers, but often need a conjunction of processes to reach the conditions for planetesimal formation. We present numerical simulations, in which the protoplanetary disk turbulence is fully captured rather than modeled with a turbulent diffusion or turbulent viscosity parameter. When the turbulent cascade is taken into account, and in the case of weakly turbulent disks, not only can solid grains be highly concentrated in clusters, but their radial drift can also be slowed or even halted. These results open a unique path to planetesimal formation starting at disk canonical dust-to-gas ratio, namely Keplerian turbulence.
\end{abstract}



\section{Introduction}
The process of planetesimal formation remains an enduring major open question in the field of planet formation research. In the incremental model of solid growth from dust to planets, the physical process allowing for the growth from grain to planetesimal continues to elude understanding. Two major barriers have been identified: the drift barrier and the collision/fragmentation barrier. With regard to the former, the gas drag exerted on the solids tends to decelerate their rotation, thus precipitating their radial drift towards the star. In the case where the timescale for this drag, $t_s$, is of the order of the rotation timescale $\Omega^{-1}$ (for a Keplerian Stokes number $t_s\Omega\sim 1$) a significant proportion of the solids may be lost to the star on a timescale too short to account for their growth. In the same range of Stokes numbers, collisions between particles result in either bouncing or fragmentation when turbulence is considered to be homogeneous and isotropic \citep{OrC07,Bir23} preventing their growth.

A number of approaches have been proposed to surmount these obstacles. To illustrate this point, consider the fluffy growth approach, which is predicated on the premise that porous solids can overcome these barriers before being compacted \citep[e.g.][]{MGP24}. The majority of current approaches are centred on the local gravitational collapse of a pebble cloud. In order to achieve a gravitationally bound state, it is 
necessary to concentrate the solids by up to four orders of magnitude, a process that typically involves several sequential steps. First, a pressure bump must be established, then drift will enable the achievement of a dust-to-gas density ratio of order one. Following this, another process, such as the streaming instability, can operate to trigger gravitational collapse \citep{YG05,SBB22}.

In this paper, the latter approach of local gravitational collapse is considered. However, a unique physical process has been identified that facilitates the concentration of solids from the typically observed low dust-to-gas ratio to levels required for gravitational collapse, while concurrently restricting the drift of solids. The objective of this study is to ascertain the impact of Keplerian turbulence on dust dynamics and to delineate the discrepancies between Keplerian turbulence and more conventional homogeneous and isotropic turbulence \citep[e.g.][]{HoC01}. The following presentation presents a novel perspective on the results of \citet{GMB22,GBM24b}, with the aim of ascertaining their astrophysical implications and addressing the challenging issue of planetesimal formation.

\section{Methods}

A concise overview of the numerical methods is provided here, with a more comprehensive and detailed presentation available in \citet{GMB22, GBM24b}.

It is noteworthy that the turbulent velocity fluctuations in protoplanetary disks are regarded as having a low amplitude, thus allowing for the safe assumption that they are subsonic. Consequently, an incompressible approach is considered at scales that are smaller than the scale height  \citep[e.g.][]{LL05, MFL15}

\begin{figure}
  \centering
 \raisebox{-0.5\height}{\includegraphics[width=0.38\textwidth,clip]{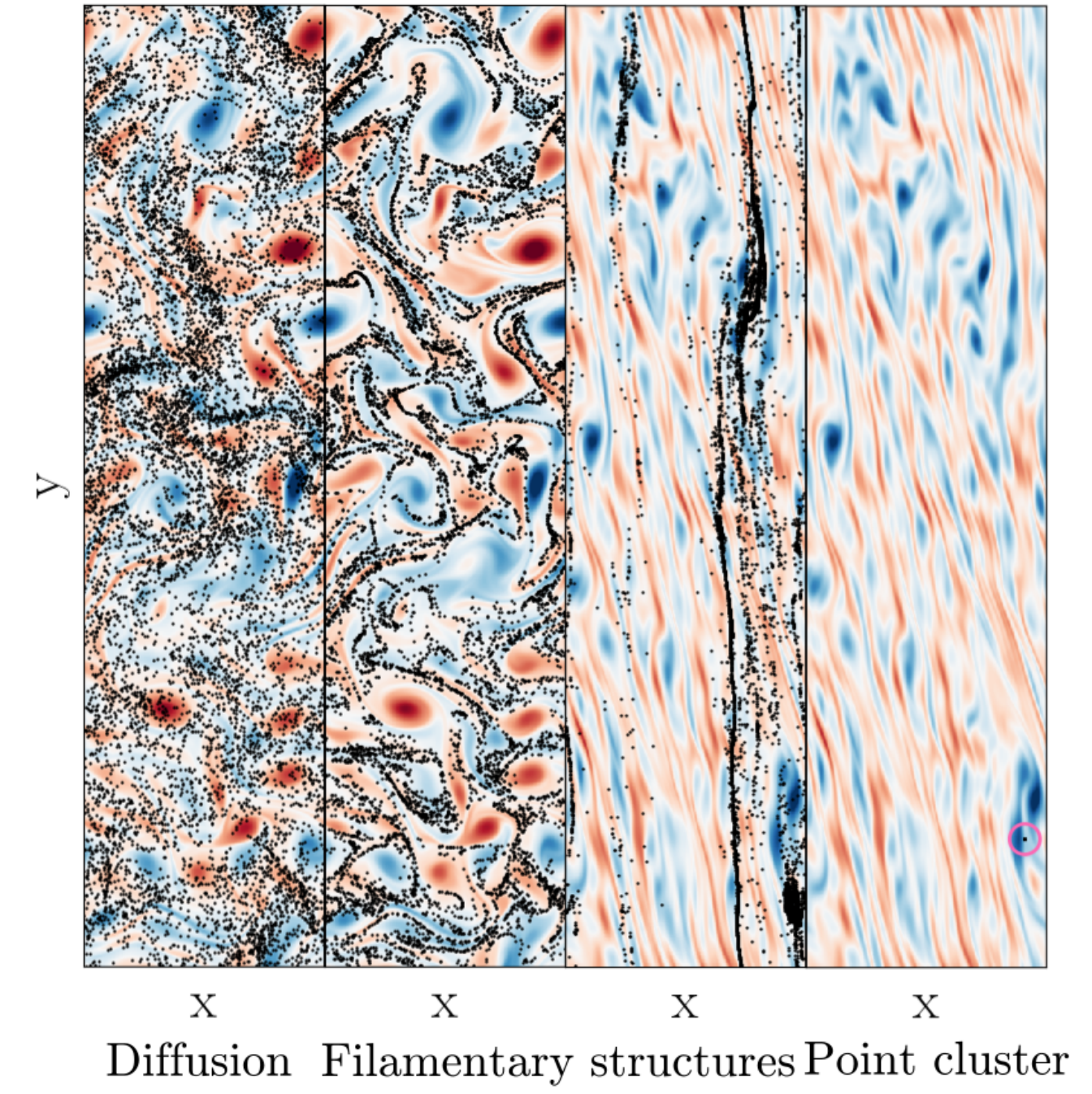}}%
 \raisebox{-0.5\height}{\includegraphics[width=0.54\textwidth,trim={0cm 0 2cm 0},clip]{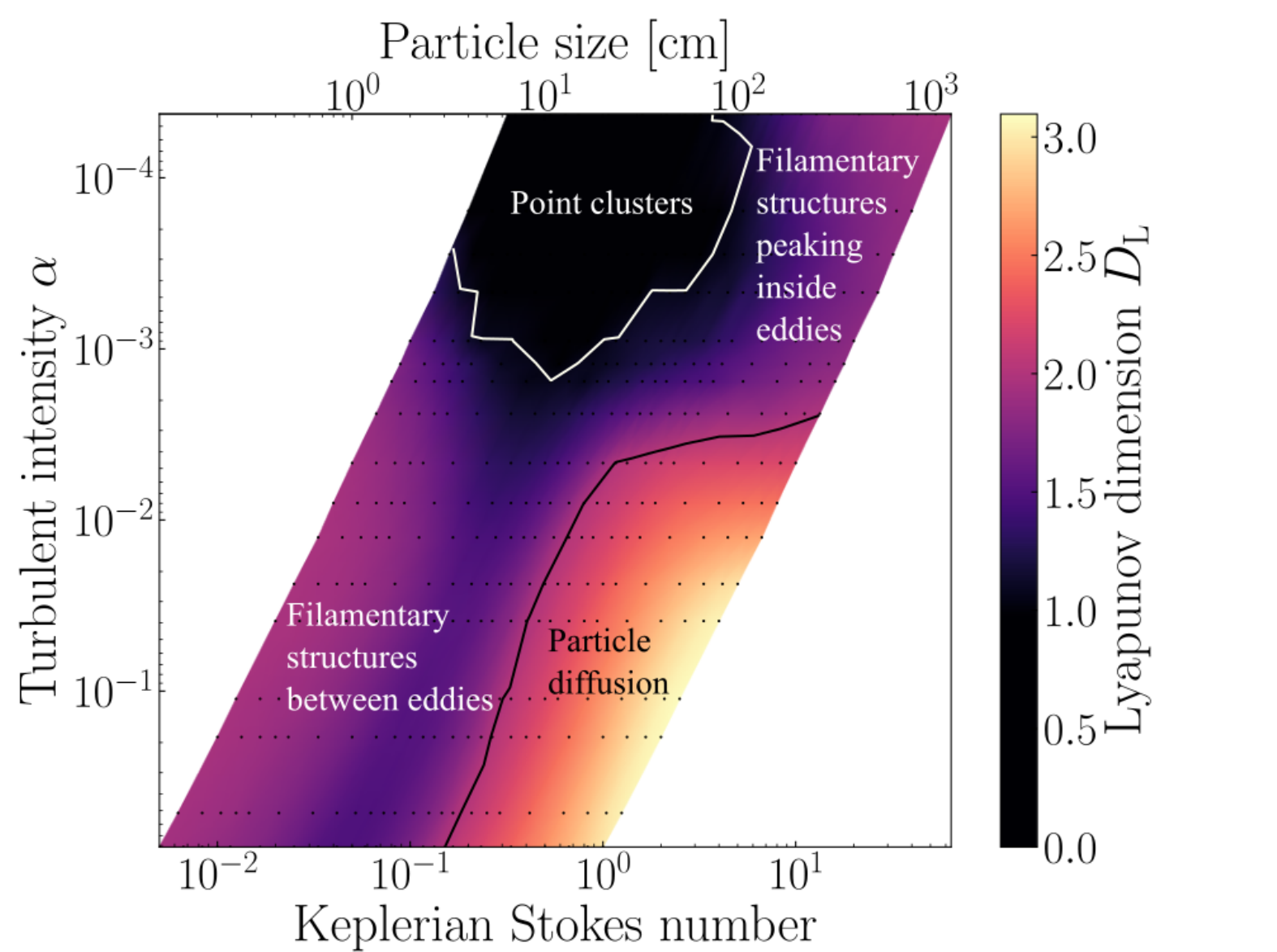}}%
  \caption{{\bf Left:} Examples of outcomes of dust structures formed in protoplanetary disk turbulence. The gas vorticity is plotted in colors, and particle positions during the statistically steady state in black. On the 4th panel, the unique particle cluster is highlighted with a circle. {\bf Right:} Summary of the different dynamical behaviors of particles in protoplanetary disk turbulence, classified according to the Lyapunov dimension of the particle set in phase space.}
  \label{Meheut:fig1}
\end{figure}

In order to study turbulence in an accretion disk, we employ the conventional shearing-box approximation \citep{HGB95} incorporating Keplerian shear and rotation. It is important to note that the onset of turbulence is not captured within the scope of the simulation. Instead, a turbulent state is sustained by imposing a large-scale external forcing at a spatial wavenumber $k_f=4$. In this study, we focus on two-dimensional turbulence that is consistent with the rotating nature of Keplerian turbulence. The implications of this simplification are discussed in the final section.

The simulations are conducted with a spectral method that exhibits low numerical dissipation, enabling the attainment of high Reynolds numbers at our moderate resolution of $[256\times 1024]$ in a box elongated in the azimuthal direction $[L\times 4L]$. For a comprehensive discussion of all methodological choices see \citet{GER24}.

The solids are treated as Lagrangian particles in interaction with the gas through linear drag (Epstein or Stokes regime). A thorough investigation has been conducted into a wide range of turbulent Rossby numbers, thereby facilitating an in-depth exploration of the entire spectrum of turbulent amplitude that can be anticipated in disks. Finally, to isolate the effect of turbulence in concentrating solids at arbitrarily low dust-to-gas ratios, the simulations neglect particles back-reaction on
the gas and discard clustering mechanisms such as the streaming instability.

\section{Results}

\subsection{Keplerian turbulence clustering}
In the primary set of simulations, the dynamics of $240,000$ Lagrangian particles are tracked, distributed uniformly across $24$ families of prescribed stopping times. These stopping times correspond to turbulent Stokes number ranging from approximately $2\times 10^{-2}$ to $9$ or, equivalently, Keplerian Stokes numbers in the range $\sim[5\times 10^{-3},60]$ \citep[see e.g.][for the two definitions of Stokes number]{SCU24}. The latter can be readily converted  (in the Epstein regime) to the corresponding particles size estimated at 5AU between $[0.1, 10^3]\mathrm{cm}$ for a gas surface density of $ 100\mathrm{g.cm^{-2}}$ and a solid density of $3 \mathrm{g.cm^{-3}}$. In this study, we explore a range of turbulent Rossby numbers \citep[see again][]{SCU24} corresponding to a range of turbulent intensity relative to disk rotation. The mean-field approach of turbulence is typically characterised by the $\alpha$ parameter, which is estimated for each simulation as the normalized mean Reynolds stress \citep[see e.g.][]{MFL15}. In order to accomplish this, it is hypothesised that the turbulent forcing length is equivalent to $ H/10 $, with $H$ representing the disk scale height.

Three different dynamical behaviors have been identified (see Fig. \ref{Meheut:fig1}): firstly, diffusion of particles dispersed throughout the domain; secondly, the formation of filamentary structures within the solid distribution, and thirdly, concentration to point clusters where all the particles of a family reach the same position with the same velocity, therefore evolving as a single particle. As illustrated on Fig. \ref{Meheut:fig1} right, such behaviors can be readily identified through the utilisation of sophisticated tools, such as the Lyapunov dimension \citep{KaY79}.

In cases where turbulent intensity is low, with a value of $\alpha$ approximating $\sim 10^{-3} - 10^{-4}$, the results demonstrate that pebble size solids ($1$cm - $1$m) are concentrated in point clusters. These point clusters are located in anticyclonic eddies (plotted in blue on Fig. \ref{Meheut:fig1} left), and can be interpreted as resulting from the concentrating effect of the Coriolis force \citep{BAR95,TBD96,CHA00}, which dominates over the expelling effect of centrifugal force, within eddies. In contrast, the prevailing interpretation of turbulent diffusion is observed under conditions of elevated turbulent intensity (extending down to $\alpha \sim 10^{-2}- 10^{-3}$), and in the size 
range of pebbles to rock-sized solids. The turbulent concentration of solids, as previously discussed by \citet{HoC01}, is observed as filamentary structures obtained in strong-amplitude turbulence (Fig. \ref{Meheut:fig1} left $2^{nd}$ panel). Finally, elongated filamentary structures, which bear a resemblance to those formed by the streaming instability, are present in low-amplitude turbulence and large particle sizes (Fig. \ref{Meheut:fig1} left $3^{rd}$ panel) without any supplementary concentration due to the back-reaction of the solids onto the gas.

\subsection{Radial drift}

\begin{figure}
 \centering
 \raisebox{-0.5\height}{\includegraphics[width=0.48\textwidth,clip]{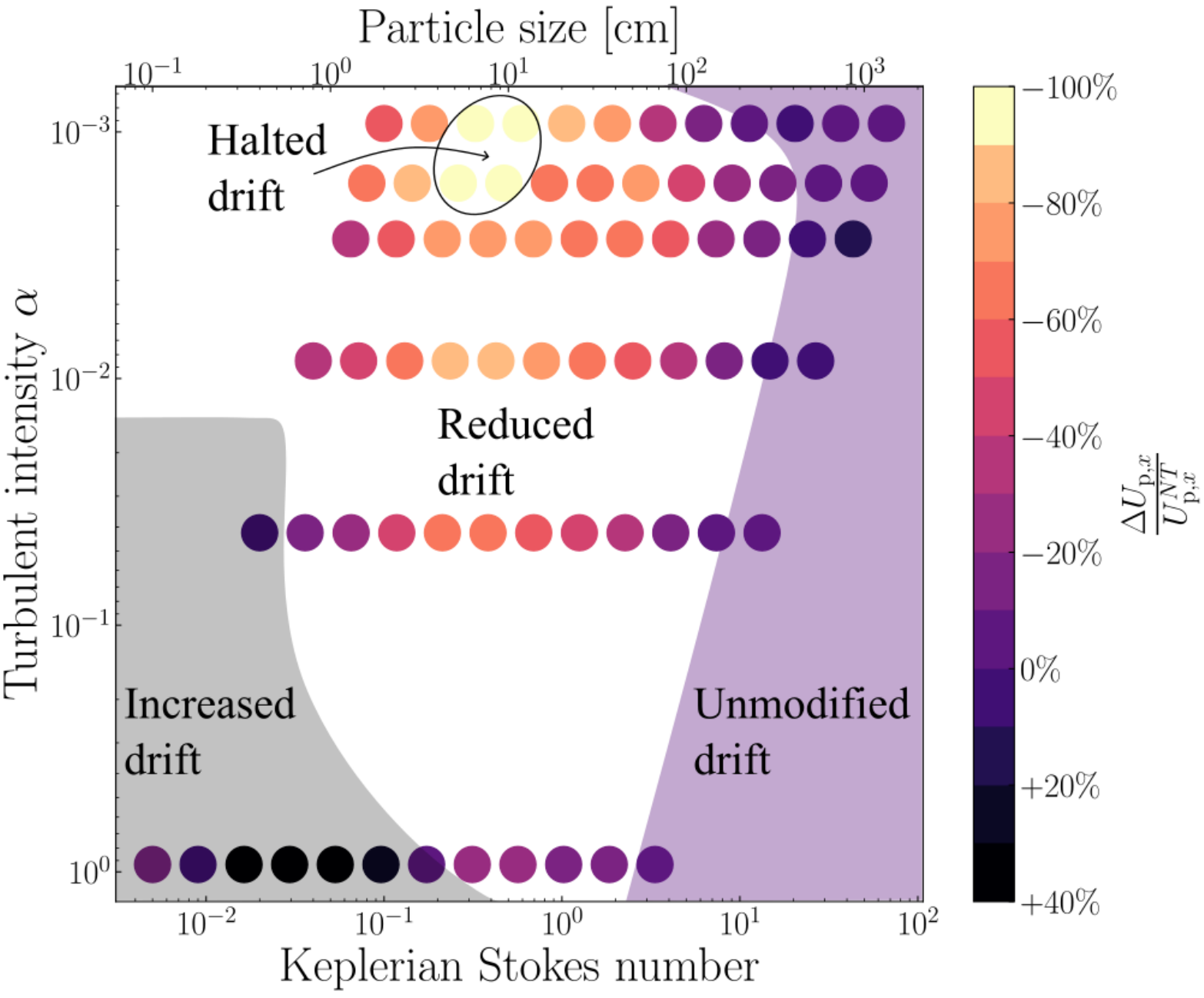}}%
 \hspace{0.5cm}
 \raisebox{-0.5\height}{\includegraphics[width=0.1\textwidth,clip]{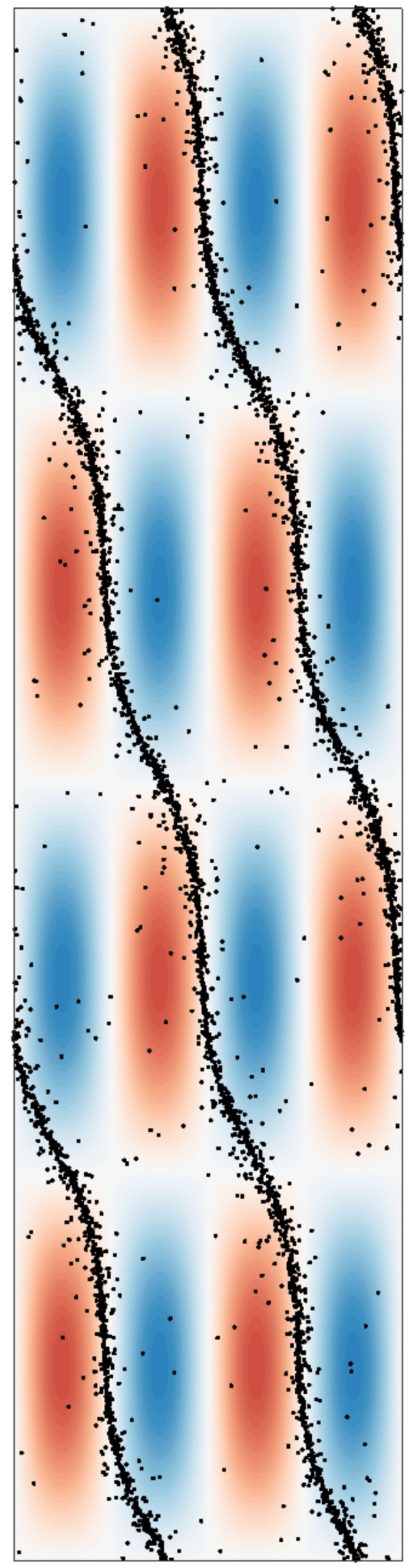}}
  \caption{{\bf Left:} Drift velocity modification by protoplanetary disk turbulence. {\bf Right:} Toy-model of particles circumventing elongated eddies resulting in reduced drift velocity. The vorticty of the flow is plotted in colors and the particles positions during a statistically steady state in black. The radial drift towards the star is oriented from right to left.}
  \label{Meheut:fig2}
\end{figure}

In a separate series of simulations, the drift of the particles was incorporated through the introduction of an additional constant azimuthal force acting on the solids.
The resulting drift velocity of the solids in the laminar case, which can be computed analytically \citep[e.g.][]{NSH86}, is compared with that obtained in the presence of turbulence. The extent of the velocity reduction or amplification is plotted in Fig.\ref{Meheut:fig2}.

It is only in circumstances where turbulence is high in intensity and particles are small in size that the conventional increase in drift velocity due to turbulence is recovered. In the major part of the parameter space, the radial drift is characterised by a reduction in velocity due to turbulence. The phenomenon can be elucidated, for $\alpha > 10^{-2}$ by considering the elongated geometry of the turbulent eddies that the particles must bypass to undergo radial drift. This unexpected behavior is evidenced by a toy model of the elongated eddies, which demonstrates that the low radial extent of the eddies limits the space over which the particles are accelerated in their drift, whereas the elongated azimuthal extent blocks the radial displacement (Fig.\ref{Meheut:fig2} right). For $\alpha < 10^{-2}$, dust mostly lies inside anticyclonic eddies, where it evolves with the gas radial velocity. In conditions of low turbulent intensity ($\alpha = 10^{-3}$), the drift of objects of sizes between centimetres and metres is reduced by a factor between 0.5 and 1.0. Furthermore, in regimes where solids form point clusters in the center of eddies, the drift can be fully arrested.

\section{Discussion and conclusion}

The present study has expanded the investigation of particle dynamics in protoplanetary disk turbulence to encompass low-amplitude turbulent intensity, characterised by a Rossby number, $Ro < 1$. To further elucidate, we define Keplerian turbulence as a fully developed turbulence wherein the effects of Keplerian rotation and shear predominate over those of turbulent vorticity. We identified Keplerian turbulence as a distinct regime relevant to planetesimal formation. It facilitates the aggregation of solids in anticyclonic eddies or filamentary structures, while concomitantly impeding or diminishing particle drift. Through this unique process, the multiple barriers to planetesimal formation can be overcome.

 It is noteworthy that the proposed approach successfully reproduces the established outcomes associated with high-amplitude turbulence, including phenomena such as particle diffusion, turbulent concentration, and an increase in drift velocity. The process identified here as \textit{Keplerian turbulence clustering} is distinct from the turbulent concentration of \citet{HoC01}, which operates at larger $Ro$ and attains lower concentrations of grains than Keplerian turbulence.
The results presented herein are restricted to a 2D approach, as rotation is well-documented to produce 2D columnar vortices in 3D simulations \citep{Tay22}. Indeed, as demonstrated by our 3D simulations, the formation of strong clusters of particles that take the form of vertical lines or surfaces, in the absence of vertical stratification, is confirmed for Keplerian turbulence (Gerosa et al., in prep).

\begin{acknowledgements}
  This work received support from the UCA-JEDI Future Investments at the Université Côte d’Azur (funded by the French government, and managed by the National Research Agency, ANR-15-IDEX-01). This work was supported by the “Programme National de Planétologie” (PNP), “Programme National de Physique Stellaire” (PNPS) of CNRS/INSU co-funded by CEA and CNES. The authors are grateful to the OPAL infrastructure and the Université Côte d’Azur’s Center for High-Performance Computing for providing resources and support.
\end{acknowledgements}


%
\end{document}